\documentclass[fleqn,12pt,twoside]{article}
\usepackage[headings]{espcrc1}

\readRCS $Id: espcrc1.tex,v 1.2 2004/02/24 11:22:11 spepping Exp $
\usepackage{graphicx}
\usepackage[figuresright]{rotating}

\newcommand{\AmS}{{\protect\the\textfont2
  A\kern-.1667em\lower.5ex\hbox{M}\kern-.125emS}}

\usepackage{amsmath}
\usepackage{amssymb}
\usepackage{epsfig}

\newcommand{\ba}{\begin{eqnarray}}
\newcommand{\ea}{\end{eqnarray}}
\newcommand{\rmi}[1]{{\mbox{\scriptsize #1}}}
\newcommand{\rmii}[1]{{\mbox{\tiny #1}}}

\newcommand{\tr}{{\rm tr\,}}

\newcommand{\fr}[2]{{\frac{#1}{#2}\,}}

\renewcommand{\(}{\left(}
\renewcommand{\)}{\right)}

\renewcommand{\ln}{{\rm ln}}

\def\half{{\textstyle \frac 12}}
\def\coeff#1#2{{\textstyle \frac {#1}{#2}}}

\def\gE{g_3}

\def\x{\mathbf {x}}
\def\Z{\mathcal{Z}}

\def\openone{\rlap 1\kern 0.22ex 1}

\newcommand{\bv}[1]{\mbox{\textbf{#1}}}

\title{Z(3)-symmetric effective theory for pure gauge QCD at high temperature}

\author{A.~Vuorinen\address[MCSD]{Department of Physics, University of Washington, Seattle, WA 98195--1560}%
        \thanks{\tt vuorinen@phys.washington.edu}}

\runtitle{Z(3)-symmetric effective theory for pure gauge QCD at high temperature}
\runauthor{A.~Vuorinen}

\begin{document}

\maketitle

\begin{abstract}
\noindent We review the construction and basic properties of a three-dimensional effective field theory for high-temperature SU(3) Yang-Mills theory, which respects
its center symmetry and was introduced in Ref.~\cite{vy}. We explain why the phase diagram of the new theory is expected to closely resemble the one of the full
theory and argue that this implies that it is applicable down to considerably lower temperatures than the usual non-Z(3)-symmetric 3d effective theory EQCD.
\end{abstract}

\section{INTRODUCTION}

Recent years have witnessed significant progress being made in the
study of equilibrium thermodynamics of high-temperature QCD on two
fronts. On one hand, the lattice QCD community, equipped with
ever-improving computational resources, has produced an abundance of
new numerical results that are valid in the region near the
deconfinement temperature $T_c$. At asymptotically high $T$, a
number of high-order perturbative calculations have on the other
hand been carried out \cite{klry}, in which a vital tool has turned
out to be the use of dimensionally reduced effective theories EQCD
and MQCD that correctly describe the behavior of static observables
in the original theory at sufficiently large length scales
\cite{bn1}. The construction of these theories relies only upon the
existence of a hierarchy between the hard, soft and ultrasoft
scales, $2\pi T\gg m_\rmii{D}\gg m_\rmi{mag}$, where $m_\rmii{D}$
stands for the electric Debye mass and $m_\rmi{mag}$ for the
magnetic mass, which is known to hold true at asymptotically high
temperatures and in addition at least to a limited extent all the
way down to $T_c$. Simulations in EQCD have, however, shown that
this theory fails to reproduce the dynamics of the full one in
the vicinity of the transition region and, in particular, that its
phase diagram has a highly unphysical structure.

At least in the case of pure Yang-Mills theory, \textit{i.e.~}$N_f=0$ QCD, the unphysical properties of EQCD at temperatures $T\sim T_c$ can
to a large extent be attributed to its failure to respect all the symmetries of the underlying theory, one of the most fundamental tenets of effective theory
building. The Yang-Mills theory has namely a global Z(3) symmetry
related to the invariance of its action under gauge transformations with twisted boundary conditions in the temporal direction, which the effective theory does not
possess. On the contrary, the leading order Lagrangian of EQCD can be obtained by expanding the one-loop effective potential of the Wilson line $\Omega(\bv x)$ in the
full theory
around one of its three degenerate Z(3)-minima, which is a valid prescription only in the close vicinity of this minimum. Thus, EQCD does not possess the
true vacuum structure of the full theory and therefore cannot correctly describe its dynamics near the deconfinement transition, which is known to be of weakly
first order and driven by tunnelings between the three deconfined Z(3)-vacua.

As a step towards obtaining an effective description of finite-temperature SU(3) Yang-Mills theory that provides a smooth interpolation between the
phase transition region and asymptotically high temperatures, it was proposed in Ref.~\cite{vy} that one should construct a three-dimensional effective theory
that respects the Z(3) symmetry, but on the other hand reduces to EQCD at high enough $T$. While reproducing all perturbative predictions obtained through
EQCD, this theory would automatically capture also the dynamics of thermal fluctuations that create bubbles sampling all three Z(3)-phases.
It should thus have a phase structure similar to that of the full four-dimensional theory, containing \textit{e.g.~}a quadruple point, where the deconfined phases
coexist with the confining one. In particular, the new theory should provide a valid description of the full theory on a temperature range
much wider than what EQCD is capable of.

In the present paper, we will briefly review the perturbative construction of a Z(3)-symmetric effective theory that was first presented in Ref.~\cite{vy}.
We will also outline the non-perturbative
simulations that need to be carried out in order to match the last undetermined parameters of the effective theory to the full one.

\section{\boldmath $Z(3)$ INVARIANT EFFECTIVE THEORY}

The minimal set of degrees of freedom in a three-dimensional Z(3) invariant effective theory for quarkless QCD consists of the spatial gauge field
$\mathbf A(\x)$ and the Wilson line $\Omega(\x)$. However, due to the unitarity of $\Omega(\x)$ and our desire to obtain a perturbatively renormalizable
theory with a polynomial Lagrangian, we choose to instead work with an unconstrained $3 \times 3$ complex matrix $\Z(\x)$, in which the extra degrees of
freedom will be chosen massive enough so that they effectively decouple from the physics of length scales $1/(gT)$ and higher. The relation between the
full theory (and the effective actions that have been built for $\Omega(\x)$ near $T_c$ \cite{pisa,kovner,peter}) and the new effective theory should be viewed as analogous
to the relation between non-linear and linear sigma models or between the Ising model and a double-well $\phi^4$ field theory.

We require that the effective theory Lagrangian be invariant under three-dimen\-sional $SU(3)$ gauge transformations
as well as global Z(3) phase rotations, and choose it to be composed of standard derivative
terms for $\mathbf A$ and $\Z$ plus a potential for the complex scalar $\Z$,
\begin{eqnarray}
    \mathcal{L}
    &=&
    \gE^{-2}
    \left\{\strut
    \half \, \tr  F_{ij}^2
    + \tr\! \left(D_i \Z^{\dagger}D_i\Z\right)
    + V(\Z)
    \right\},
\label{lageff2}
\end{eqnarray}
with $D_i \equiv \partial_i -i [A_i,\,\cdot\,]$
and $F_{ij} \equiv \partial_i A_j - \partial_j A_i - [A_i,A_j]$.
Defining
\ba
    L(\x) &\equiv& \tr \Z(\x), \quad M(\x) \equiv \Z(\x) -\coeff{1}{3} \, \tr \Z(\x)\,\openone \,,
\ea
the $Z(3)$ invariant potential $V(\Z)$ will consist of two pieces,
\ba
    V(\Z) &=& V_0(\Z) + \gE^2 \, V_1(\Z) \,,
\label{eq:V}
\ea
with
\begin{align}
    V_0(\Z)
    &\;=\;  
    c_1 \,\tr\!\big[\Z^{\dagger}\Z\big]
    + c_2 \({\rm det}\!\big[\Z\big] + {\rm det}\big[\Z^{\dagger}\big]\)
    + c_3\, \tr\!\left[ (\Z^{\dagger}\Z )^{2}\right] ,
\label{v0}\\
    V_1(\Z)
    &\;=\;  
    \tilde{c}_1 \, \tr\!\big[M^{\dagger}M\big]
    + \tilde{c}_2 \(\tr [M^3] + \tr [(M^{\dagger})^{3}]\)
    + \tilde{c}_3 \, \tr\!\big[(M^{\dagger}M)^2\big] \,.
\label{v1}
\end{align}
The first of these is referred to as the ``hard potential''
and the second as the ``soft potential'', since at high temperatures and small $\gE^2/T$, $V_0(\Z)$ will give the heavy excitations their $\mathcal O(T)$ masses,
while $\gE^2 \, V_1(\Z)$ will be responsible for the $\mathcal O(gT)$ masses of the physical components of $\Z$. The theory is a superrenormalizable,
as the only renormalizations required (in addition to vacuum energy subtraction)
are one- and two-loop adjustments of the coefficients $c_1$ and $\tilde c_1$ of the quadratic terms.

The above choices for the potentials are such that for $c_2 < 0$ and $c_2^2 > 9\, c_1 c_3$, $V_0$ is minimized by
an arbitrary $SU(3)$ matrix times $v/3$, with
\ba
    v
    &\equiv&
    \coeff 34
    \biggl(
    \frac{-c_2 + \sqrt{c_2^2 - 8 c_1 c_3} }{c_3} \>
    \biggr) ,
\label{eq:v}
\ea
while for $\tilde c_3 > 0$ and $\tilde{c}_1\tilde{c}_3>\tilde{c}_2^2$, $V_1$ is minimized by $M=0$. Thus, with these parameter values
the entire potential $V$ is minimized by
$\Z=\coeff v3 {\rm e}^{2\pi i n/3}\openone$, with $n\in\{0,1,2\}$, corresponding to the three $Z(3)$ minima of the effective potential for $\Omega$ in the
full theory. While the form of $V_0$ was dictated by computational convenience (see Ref.~\cite{vy}), the choice of $V_1$ was to some extent
arbitrary, as we could have added to it several other $Z(3)$ invariant terms with the same dimensions. However, for $Z$ restricted to the minima of $V_0$,
there are only three independent operators which are gauge invariant, $Z(3)$ invariant, and at most fourth order in the field, for which the three terms
we have included in the potential of Eq.~(\ref{v1}) are
a convenient choice.

The parameters appearing in Eqs.~(\ref{v0})--(\ref{v1}) are available through straightforward calculations in the effective theory,
once we demand that the following conditions are met:
\begin{itemize}
\item The integration-out of the heavy degrees of freedom in the effective theory produces a theory for the spatial gauge field $\mathbf A$ and
a traceless hermitian scalar field $a$, whose leading order Lagrangian density agrees with that of EQCD with the identification $a\leftrightarrow A_0$.
\item The leading order tension and width of the effective theory domain wall separating two distinct $Z(3)$ minima agree with the corresponding full theory results.
\end{itemize}
It was shown in Ref.~\cite{vy} that to leading order this leads to the results
\begin{eqnarray}
    \tilde{c}_1&=&T,\quad
    \tilde{c}_2 \;=\; 0.118914,\quad
    \tilde{c}_3\;=\;\fr 3{4 \pi^2 T},\quad
    v\;=\;3.005868T,
\end{eqnarray}
leaving only two dimensionless numbers corresponding to the ratios of the masses of the heavy excitations undetermined.

\section{PHASE DIAGRAM AND NON-PERTURBATIVE TESTS}

At high temperatures, \textit{i.e.~}deep in the deconfined phase, the perturbative matching performed in Ref.~\cite{vy}
guarantees that new effective theory reproduce the correct long-distance
physics of the full one, but close to $T_c$ we need additional non-perturbative matching.
Keeping the heavy masses as free parameters for now, we can, however, already make some generic remarks about the properties of the new theory.
Tuning the values of these
quantities, one can shift the minima of the tree level potential from the three deconfined Z(3) minima to a confining one at $\Z=0$, which corresponds
to a strongly first order phase transition at weak coupling. Increasing the value of $\gE$, the fluctuations around the minima are on the other hand enhanced,
which we believe will turn the transition to weakly first order. Thus, we expect our theory to be able to mimic the phase transition of the full theory to at least
some extent, once its parameters are properly adjusted.

To make the above statements more quantitative and, in particular, to find optimal values for the undetermined parameters, with which the correlation lengths
of various observables agree with the full theory ones even near the phase transition, it is necessary to perform non-perturbative lattice simulations
with the effective theory Lagrangian. There, the first task should be the mapping of the
phase diagram of the new theory in terms of $\gE^2/T$ and the mass ratios, and in particular to the determination of the order of the phase transition line.
Work towards this goal is currectly being carried out \cite{Aleksi2}.

\begin{figure}[t]
\epsfig{file=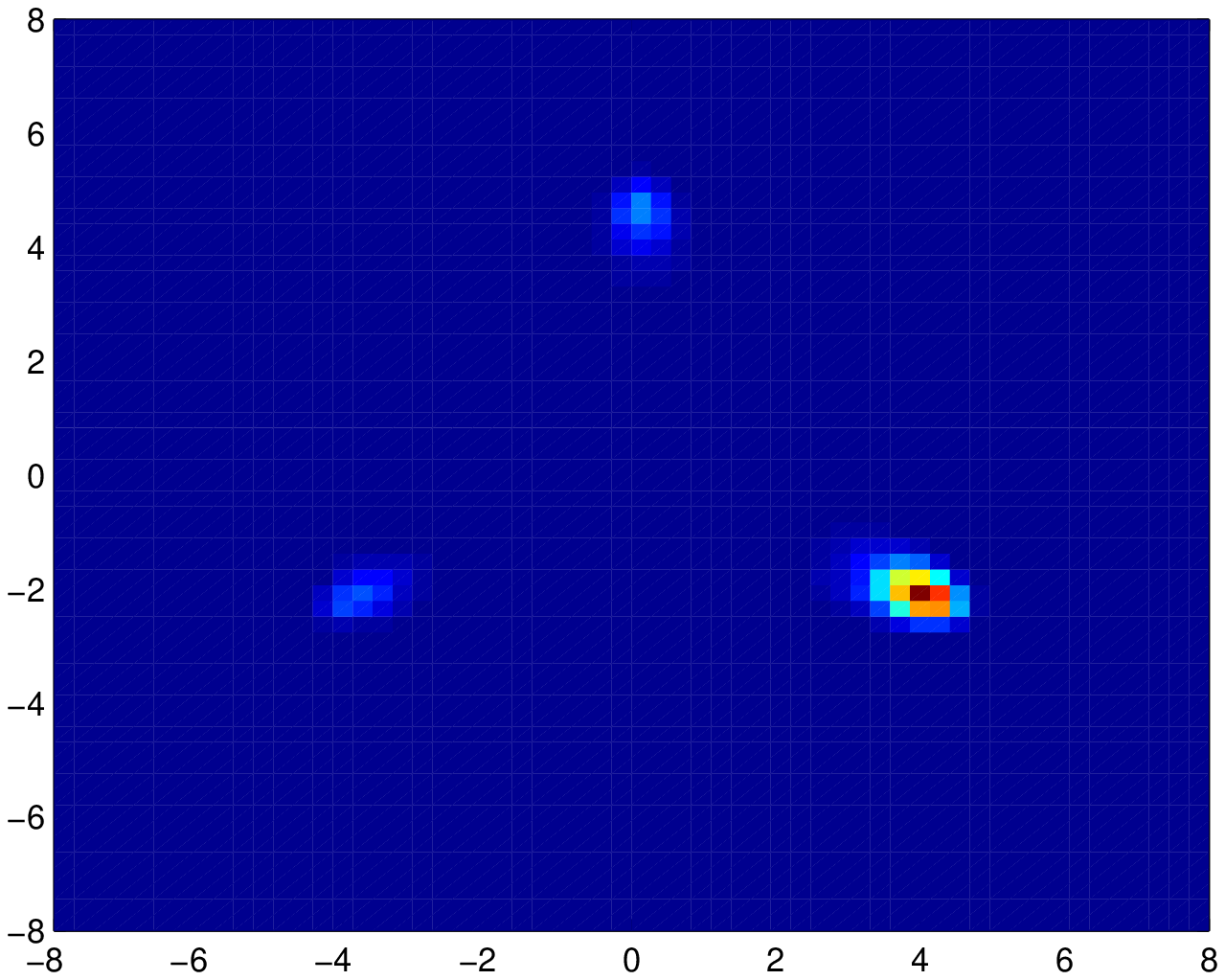,width=4.4cm,height=4.4cm}\hspace{-0.7cm}
\epsfig{file=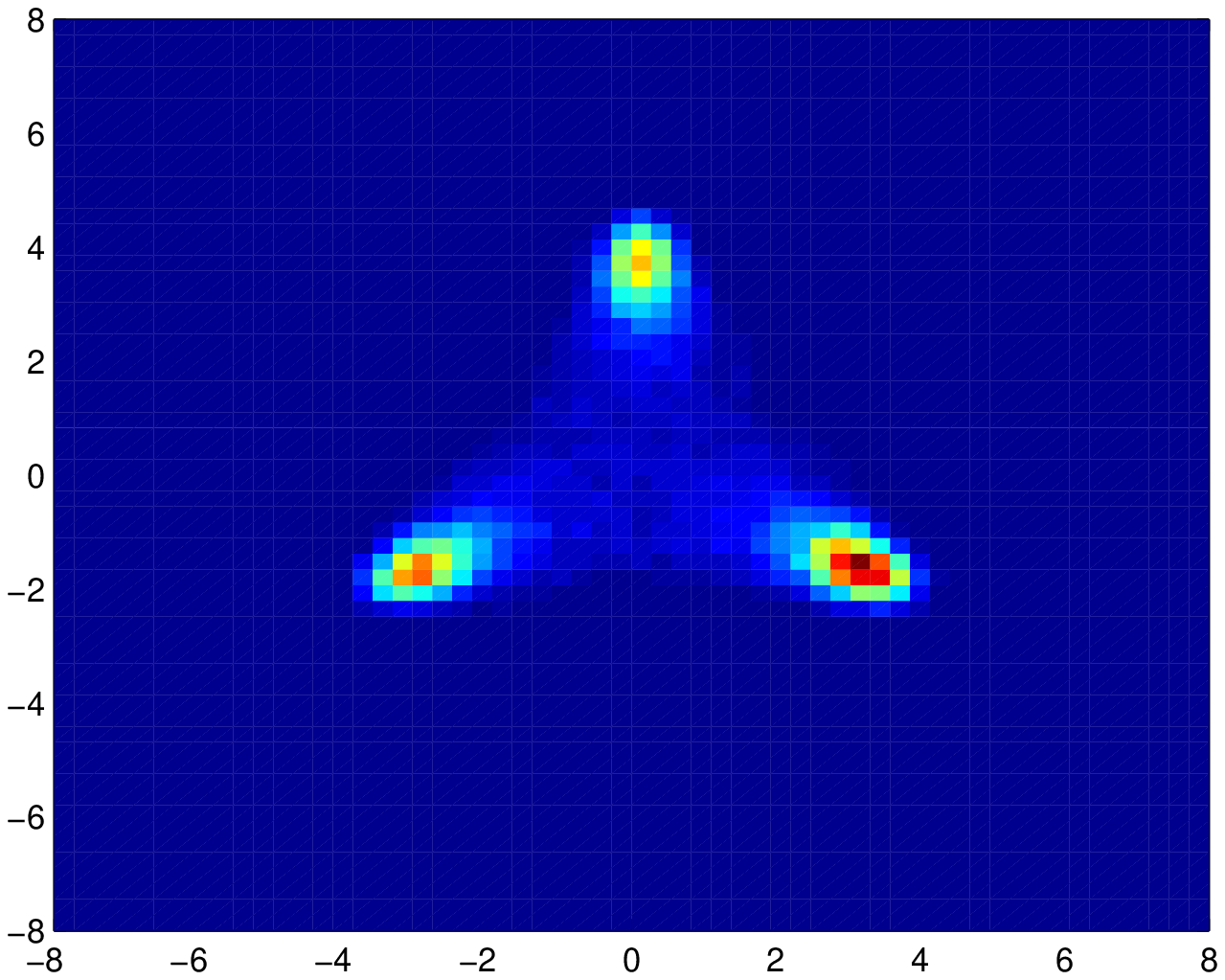,width=4.4cm,height=4.4cm}\hspace{-0.7cm}
\epsfig{file=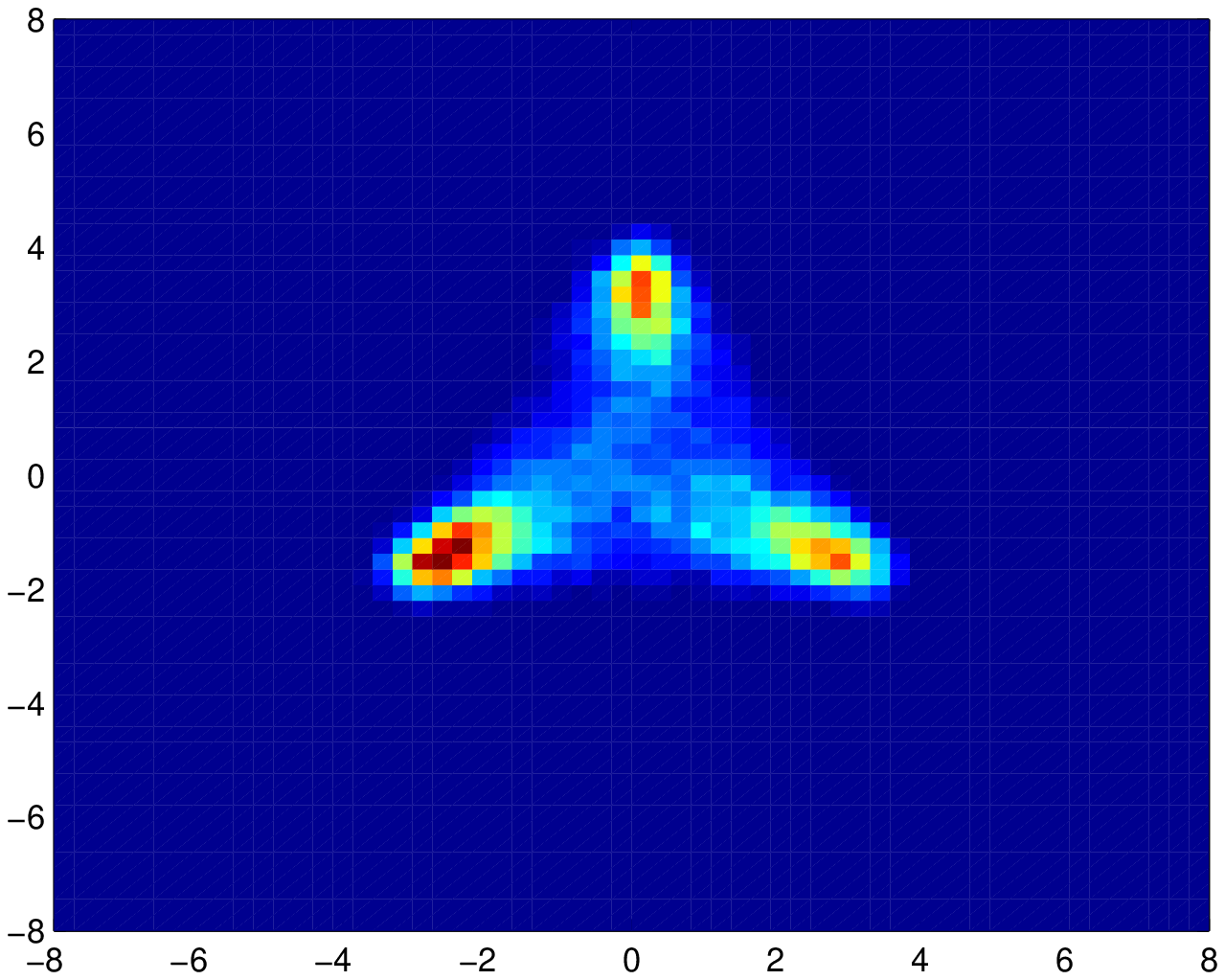,width=4.4cm,height=4.4cm}\hspace{-0.7cm}
\epsfig{file=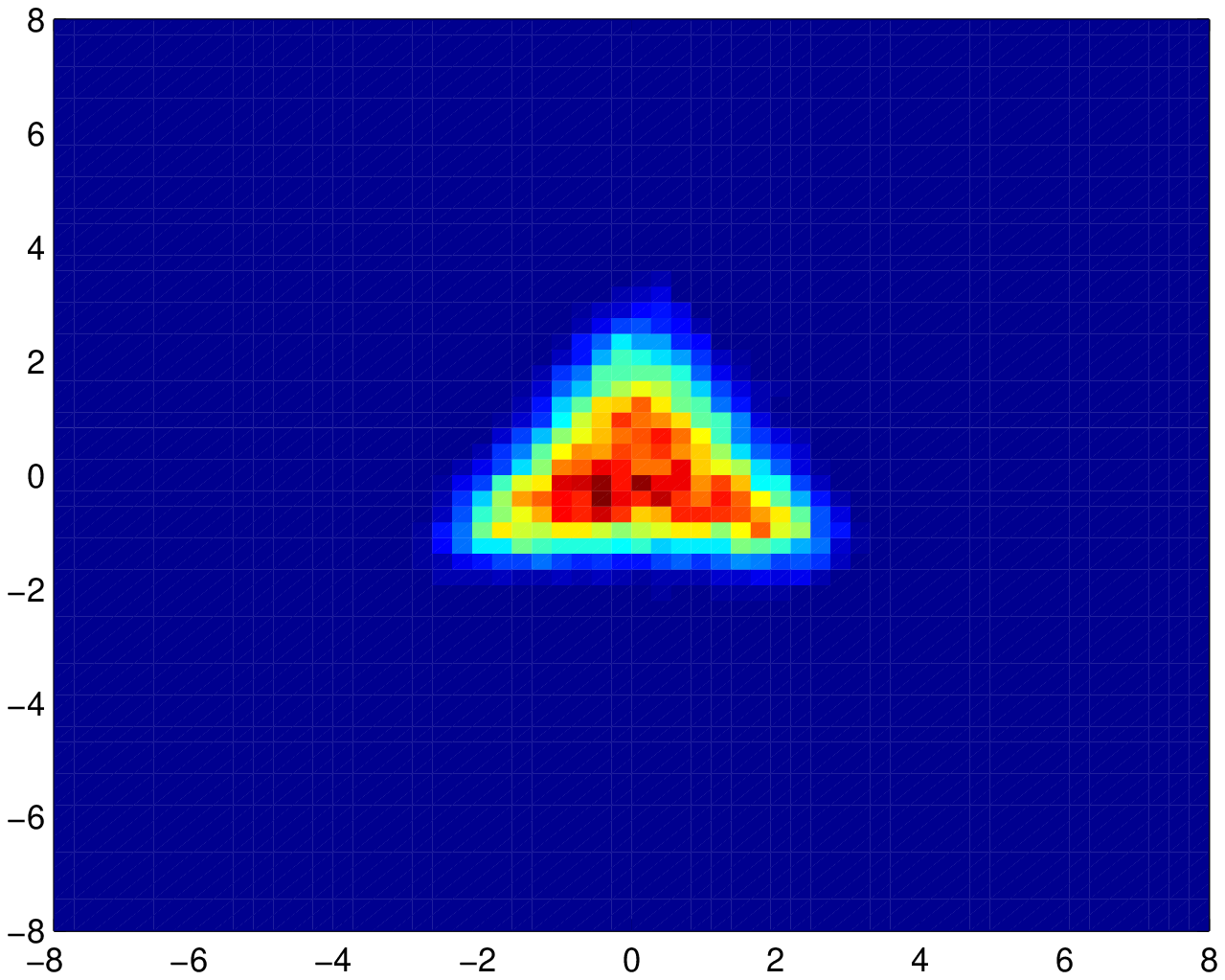,width=4.4cm,height=4.4cm}
\caption{Preliminary results for the probability density distribution of $\tr\Z$ on a (rotated) complex plane \cite{Aleksi2}. The value of $\gE^2/T$ is increased from left to
right but the heavy mass ratios are kept fixed, which makes the transition from the deconfined to the confining phase clearly visible.}
\label{fig}
\end{figure}

\end{document}